\documentclass[aps,prb,twocolumn,superscriptaddress,showpacs]{revtex4}

\usepackage{graphicx}
\usepackage{dcolumn}
\usepackage{bm}
\usepackage{Jonasmacros}

\def\kpar{\mathbf{k}_\parallel}

\begin{document}

\title{Detection of the spin character of Fe(001) surface states by Scanning Tunneling Microscopy: A theoretical proposal}

\author{Athanasios N. Chantis}
\thanks{achantis@lanl.gov}
\affiliation{Theoretical Division, Los Alamos National Laboratory, Los Alamos, New Mexico 87545, USA}
\author{Darryl L. Smith}
\affiliation{Theoretical Division, Los Alamos National Laboratory, Los Alamos, New Mexico 87545, USA}
\author{J. Fransson}
\affiliation{Department of Physics and Materials Science, Uppsala University, Box 530, SE-751 21\ \ Uppsala, Sweden.}
\author{A. V. Balatsky}
\affiliation{Theoretical Division, Los Alamos National Laboratory, Los Alamos, New Mexico 87545, USA}
\affiliation{Center for Integrated Nanotechnology, Los Alamos National Laboratory, Los Alamos, New Mexico 87545, USA}
\date{\today}

\begin{abstract}
We consider the magnetic structure on the Fe(001) surface and 
theoretically study the scanning tunneling spectroscopy using a spin-polarized tip (SP-STM). 
We show that minority-spin surface states induce a strong bias dependence 
of the tunneling differential conductance which largely depends on the
orientation of the magnetization in the SP-STM tip relative to the
easy magnetization axis in the Fe(001) surface. We propose
to use this effect in order to determine the spin character of the 
Fe(001) surface states. This technique can be applied also to other magnetic surfaces
in which surface states are observed.
 
\end{abstract}

\pacs{72.25.Mk, 73.23.-b, 73.40.Gk, 73.40.Rw}

\maketitle

\section{INTRODUCTION}

The properties of magnetic surfaces and interfaces 
have attracted recently a lot of attention because 
of the advent of spintronics,  a
technology aiming to harness electron's spin in data storage and
processing, typically by utilizing heterostructures composed of
magnetic and non-magnetic materials \cite{zutic}.
Recent theoretical studies revealed that 
the electronic properties of the Fe(001) surface can play 
a much more important role for ferromagnet-semiconductor based spintronic devices 
than it was previously thought \cite{chantis1,chantis2,khan}.
Using first-principles electron transport methods it was shown that Fe 3{\it d} minority-spin
surface (interface) states are responsible for at least two important effects for 
spin electronics. 
First, they can produce a \emph{sizable} Tunneling Anisotropic 
Magnetoresistance (TAMR) in magnetic tunnel junctions with a \emph{single} Fe electrode~\cite{chantis1}. 
The effect is driven by a Rashba shift of the resonant surface band when 
the magnetization changes direction. This can introduce a new class
of spintronic devices, namely, Tunneling Magnetoresistance junctions with a
single ferromagnetic electrode~\cite{gould} that can function at room temperatures.
Second, in Fe/GaAs(001) magnetic tunnel junctions minority-spin interface states produce
a \emph{strong} dependence of the tunneling current spin-polarization on
applied electrical bias. A 
dramatic \emph{sign reversal} within a voltage range of just a few tenths of an eV
was predicted. 
This explains the observed sign reversal of spin-polarization in 
recent experiments of electrical spin injection in Fe/GaAs(001) \cite{louNP}
and related reversal of tunneling magnetoresistcance through vertical 
Fe/GaAs/Fe trilayers. The TAMR effect was also observed recently in
a Fe/GaAs/Au tunnel junction \cite{moser}.

Many of the theoretical results mentioned above, are based on 
theoretical predictions that the Fe(001) surface band is of \emph{minority spin}.
However, to this date a \emph{direct} experimental determination of the spin 
character of the Fe(001) surface band has yet to be done. 
Scanning tunneling microscopy (STM) is a well established technique for 
imaging surface structures \cite{binnig1981,binnig1982,binnig1983,binnig_ss_1983} and 
for spectroscopic measurements of local structures and inhomogeneities on surfaces 
\cite{crommie1993,hirjibehedin2006,hirjibehedin2007}. Using this technique
Stroscio \emph{et al} provided the first experimental evidence that a surface band
near the Fermi energy exists in the Fe(001) surface \cite{stroscio}, 
however the technique used was incapable of distinguishing the \emph{spin character}
of this band.  Recent developments of STM tools with spin-polarized tip, SP-STM, 
allow for controllable measurements of 
magnetic \cite{heinze2000,wachowiak2002,bode2007} and spin-dynamical \cite{meier2008} 
features and local magnetic structures \cite{gambardella2003}. 
Spin-polarized tunneling for STM set-up has been theoretically studied with 
respect to noise \cite{nussinov2003}, 
and spin-detection and spin-reversal of local spins located on a 
substrate surface \cite{fransson2008-1,fransson2008-2}. 

In this paper, we consider a SP-STM set-up on top of a planar Fe surface 
in order to study the influence of the minority-spin resonant state at the 
surface on the spin-polarized tunneling current. We found that a
complicated angle and energy dependence of the tunneling differential conductance 
emerges as a result of the energy and momentum dependence of minority-spin 
band structure in Fe(001) surface.
Our results identify 
a specific route on how to determine the spin character of the Fe(001) 
surface band with the help of SP-STM. In principle, these results
are applicable to a broad set of materials where the minority/majority
spin-structure exhibits nontrivial energy dependence.

The rest of this article is organized as follows. In section II
we discuss our numerical approach and present our results. Particular emphasis is 
given to the relation of our approach to Tersoff and Hamman formula
for the tunneling current in STM~\cite{tersoff85}. 
Then, based on the obtained results we discuss our proposal on how to detect
the spin character of surface states with the help of SP-STM. 
Section III is the conclusion.
 
\section{MAIN}

We consider a Fe/vaccum/Cu tunnel structure with a nonmagnetic bcc Cu
electrode. This electrode has a spin-independent
free-electron-like band structure and a featureless surface
transmission function \cite{stf}, which makes it insensitive
to the transverse wavevector. This electrode simulates an ideal 
STM tip.  The structure investigated
consists of a semi-infinite Fe region, several layers of vacuum (empty atomic spheres), \
and a semi-infinite Cu region.  
Another advantage of such setup is that it avoids overall the possible 
'handshake' of surface resonances at two opposite metallic surfaces/interfaces~\cite{khan}.
When two resonant states on opposite contacts are located in 
the same $(E,\kpar)$ space then naturally a resonant transmission equal to 1 will occur
across the structure. Obviously a situation like this
may occur in a symmetric structure.
However, in real structures such 'handshakes' are unlikely because the symmetry of 
the structure is broken by applied bias or disorder. Therefore, a Fe(001)/vacuum/Fe(001)
setup may yield unphysical high transmission for surface states.

\begin{figure}[tbp]
\includegraphics[width=0.41\textwidth]{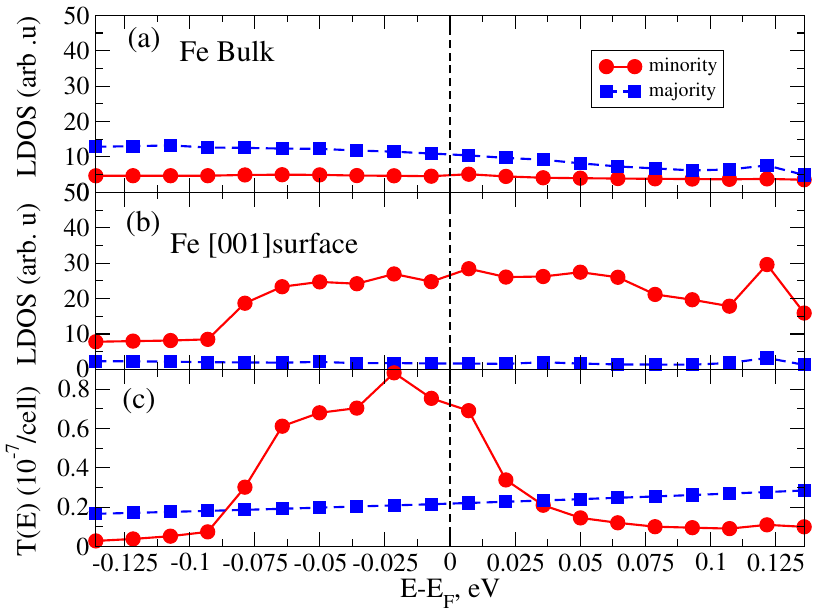}
\caption{ \small 
(a) Spin-resolved DOS for the bulk Fe 
(b) Spin-resolved DOS for the Fe(001) surface
(c) Spin-resolved $\kpar$-integrated transmission for the
Fe/vaccum/Cu as a function of energy. The Fermi level is at zero
energy.}
\label{Fig1}
\end{figure}

The calculation approach is based on the Green's function
representation of the Tight-Binding Linear Muffin-Tin Orbital
(TB-LMTO) method in the Atomic Sphere Approximation (ASA)
\cite{andersen}. We use third order parametrization for the
Green's function \cite{Gunnarson}. The electronic structure
problem is solved within the scalar relativistic Density
Functional Theory (DFT) where the exchange and correlation
potential is treated in the Local Spin Density Approximation
(LSDA).  The conductance is calculated with the principal-layer
Green's function technique \cite{Turek,kudr00,chantis3} within the
Landauer-B\"uttiker approach \cite{Datta}. 
The semi-infinite Fe and Cu electrodes are separated by
approximately 1~nm of vacuum represented by 6 monolayers of empty
spheres.   The structure is oriented
in the [001] direction. Self-consistent charge distribution is achieved
before any transport calculations are attempted. 
The spin-dependent
$\kpar$-integrated transmission 
\begin{equation}
\rm{T^\sigma(E)=1/2\pi\int_{2DBZ} t^\sigma(E,\kpar)d^2\kpar} 
\label{eq1}
\end{equation}
is calculated in the
window from $E_F$ to $E_F+$eV. $\rm{t^\sigma(E,\kpar)}$ is the
transmission coefficient and $\rm{\sigma=\uparrow,\downarrow}$
($\uparrow$=majority-spin, $\downarrow$=minority-spin).  The spin
quantization axis lies along (001) direction. A uniform
250$\times$250 mesh was used for the integration in the
two-dimensional Brillouin zone (2DBZ).  With this transmission we
can associate a current density
\begin{equation}
J^{\sigma}(V)=e/h\int_{E_{F}}^{E_{F}+eV} T^\sigma(E) dE
\label{eq2}
\end{equation},
this is an excellent approximation appropriate for comparison with
experiment when the considered voltages are
small. Then the spin resolved differential conductance is $dJ^{\sigma}/dV
\propto T(E_{F}+eV)$.



\begin{figure}[tbp]
\includegraphics[width=0.5\textwidth]{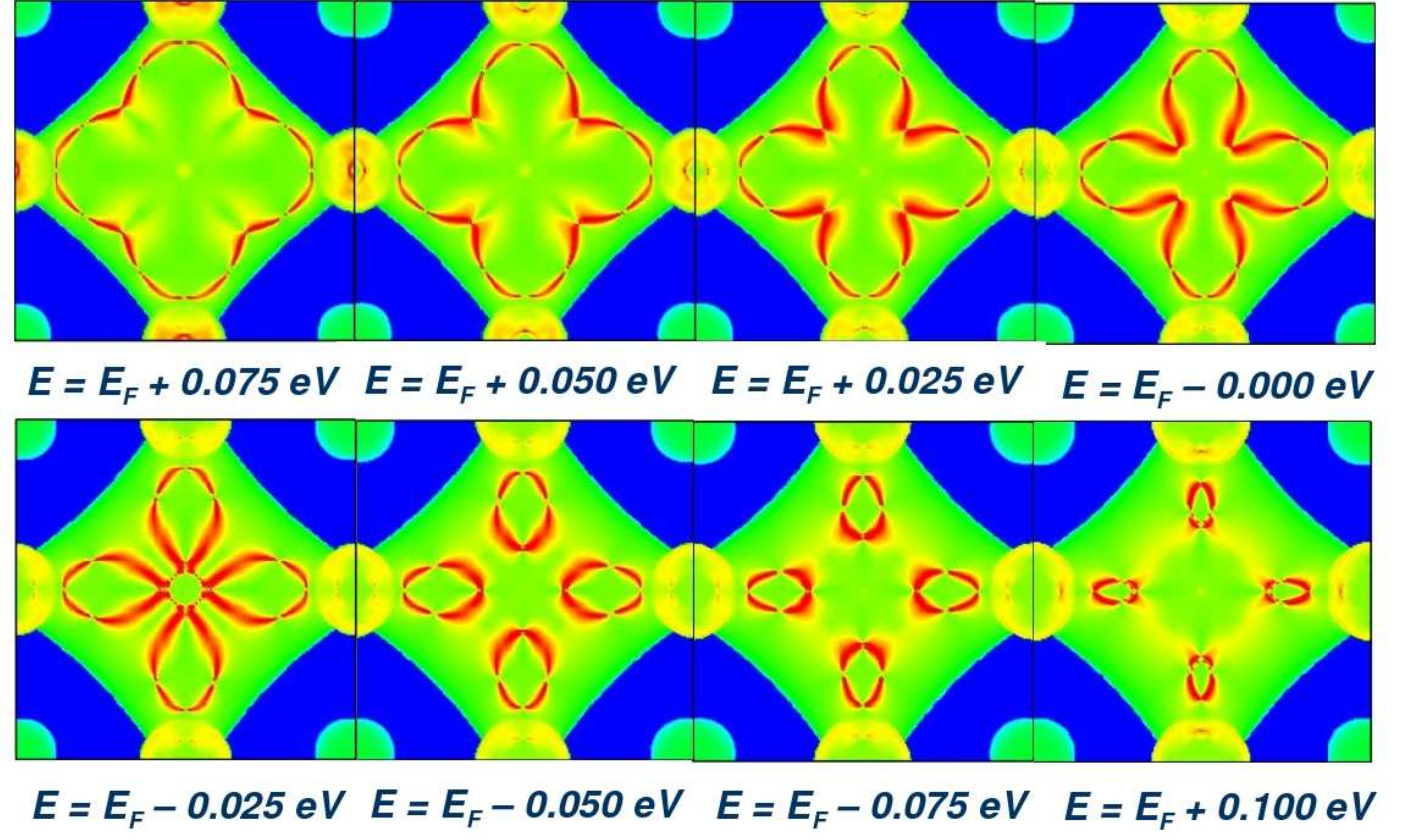}
\caption{ \small  \small Minority-spin $\kpar$-resolved DOS of the Fe(001) surface for 
different energies around the Fermi level. 
The abscissa is along [100] and the ordinate is along [010]. 
The maximum value is represented by red color, the minimum by blue.
}
\label{Fig2}
\end{figure}

Figs. \ref{Fig1}(a) and \ref{Fig1}(b)  show the calculated Fe bulk spin-resolved density of states (DOS) 
and Local DOS of the Fe monolayer at the Fe(001) surface, respectively.
The energies are given with respect to the Fermi level ${E_F}$.  In the Fe bulk (Fig. \ref{Fig1}(a)) 
the majority-spin dominates over the minority-spin
throughout the entire energy interval shown here.  However, in
the surface monolayer (Fig. \ref{Fig1}(b)) the spin polarization of the DOS is totally reversed; 
it is the \emph{minority-spin} that dominates over the majority-spin 
throughout the entire energy interval. In Refs.~\onlinecite{chantis1} and ~\onlinecite{chantis2} it was shown that this reversal
is caused by Fe $3d$ surface states of minority-spin.
As shown in Fig.~\ref{Fig1}(c), sign reversal of the spin polarization of the surface DOS
doesn't necessarily lead to sign reversal of the spin polarization of tunneling transmission, at least throughout the
same energy interval. 

\begin{figure}[tbp]
\includegraphics[angle=0,width=0.5\textwidth,clip]{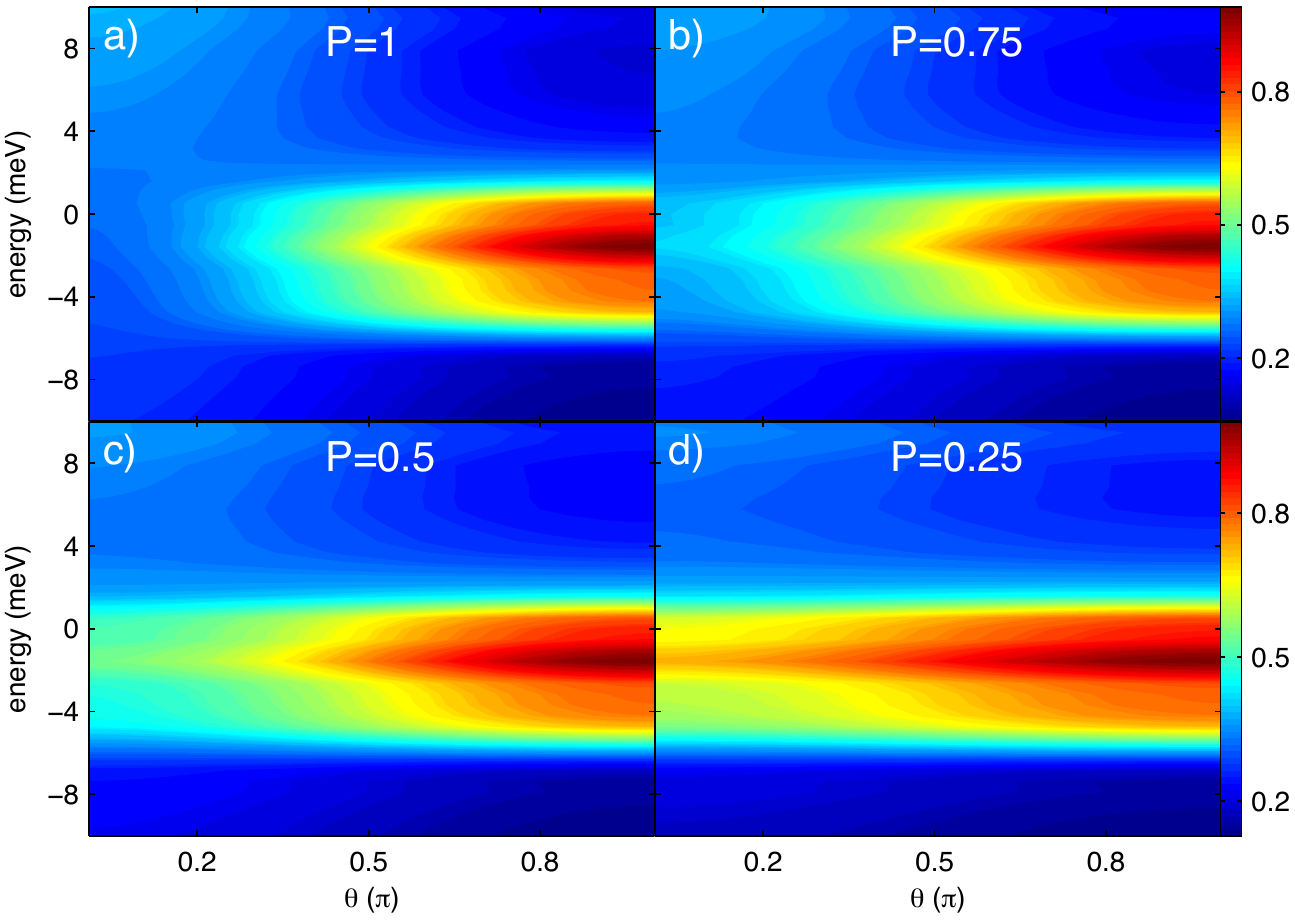}
\caption{The energy dependence of the differential conductance of a 
FM tip as we rotate the direction of the magnetization in the tip by an angle $\theta$ 
relatively to the magnetization in Fe. We show four different 
cases of DOS spin polarization $P$ of the tip. The bright red regions
observed for $\theta=\pi$ are created by the minority spin transmission
peaks in Fig.~1(c).} 
\label{Fig3}
\end{figure}

To explain this in Fig.~\ref{Fig2} we present the minority-spin $\kpar$-resolved DOS of the Fe(001) surface 
for different energies around the Fermi level. The bright red features on these plots are created 
by the surface band.  The surface band has $C_{4v}$ symmetry which is the symmetry of the
Fe(001) surface.  These bands are dominated by the
minority-spin surface states arising from $d_{x^2-y^2}$ and
$d_{xy}$ orbitals on surface Fe sites that couple with the bulk
Fe $\Delta_{2^\prime}$ minority band.
Unlike the
majority-spin bulk band this band never crosses the 2DBZ at the $\Gamma$ point. 
The closer it gets to the $\Gamma$ point is at the energy of $E_{F}-0.025 eV$ where we see
a bright four-petal structure centered at the $\Gamma$ point without touching it. 
Considering that $\kpar$ is conserved  during tunneling across an ideal surface, 
for an electron entering the vacuum region with a real $\kpar$ the decay rate of the electron wave function
is proportional to $\rm{exp\left[-\left(\kappa^2+\kpar^2\right)z\right]}$; where $\rm{\kappa}$ is
the decay rate for normal incidence which is determined by the potential height of the
tunneling barrier. 
The tunneling transmission for a given energy is also proportional to the
total number of states at this energy $n_{\sigma}(E)$. In Fig.~\ref{Fig1}(c) one can compare the spin-resolved
tunneling transmission with the spin-resolved surface DOS. We see the influence of both factors just mentioned above in 
the polarization of the transmission coefficient. 
The minority-spin transmission dominates over the majority-spin for a large part of the energy interval
due to its much higher surface DOS. For the energy $E=E_{F}-0.025 eV$, 
the minority-spin transmission has a maximum and in the energy interval where the 
minority-spin DOS is flat, the minority-spin transmission is less when the surface states
are further away from the $\Gamma$ point.    
For energies where the minority-spin surface states are far away from the center of the 2DBZ, the minority-spin
transmission is less than the majority-spin even though for the same energy 
the minority-spin surface DOS is much larger than the majority-spin.  
For bigger distances between the Cu counterelectrode and the Fe surface
the ratio of minority-spin to majority-spin transmission should become even less
for all energies in the shown interval. 
However, our calculations show that for a distance twice as big, the change of spin polarization bias
dependence is not very big while the current drops by about four orders of magnitude. 
This shows that within the range of distances 
available to an SP-STM tip, the inversion of spin polarization should be detectable
(below we describe how). 


\begin{figure}[tbp]
\includegraphics[angle=0,width=0.5\textwidth,clip]{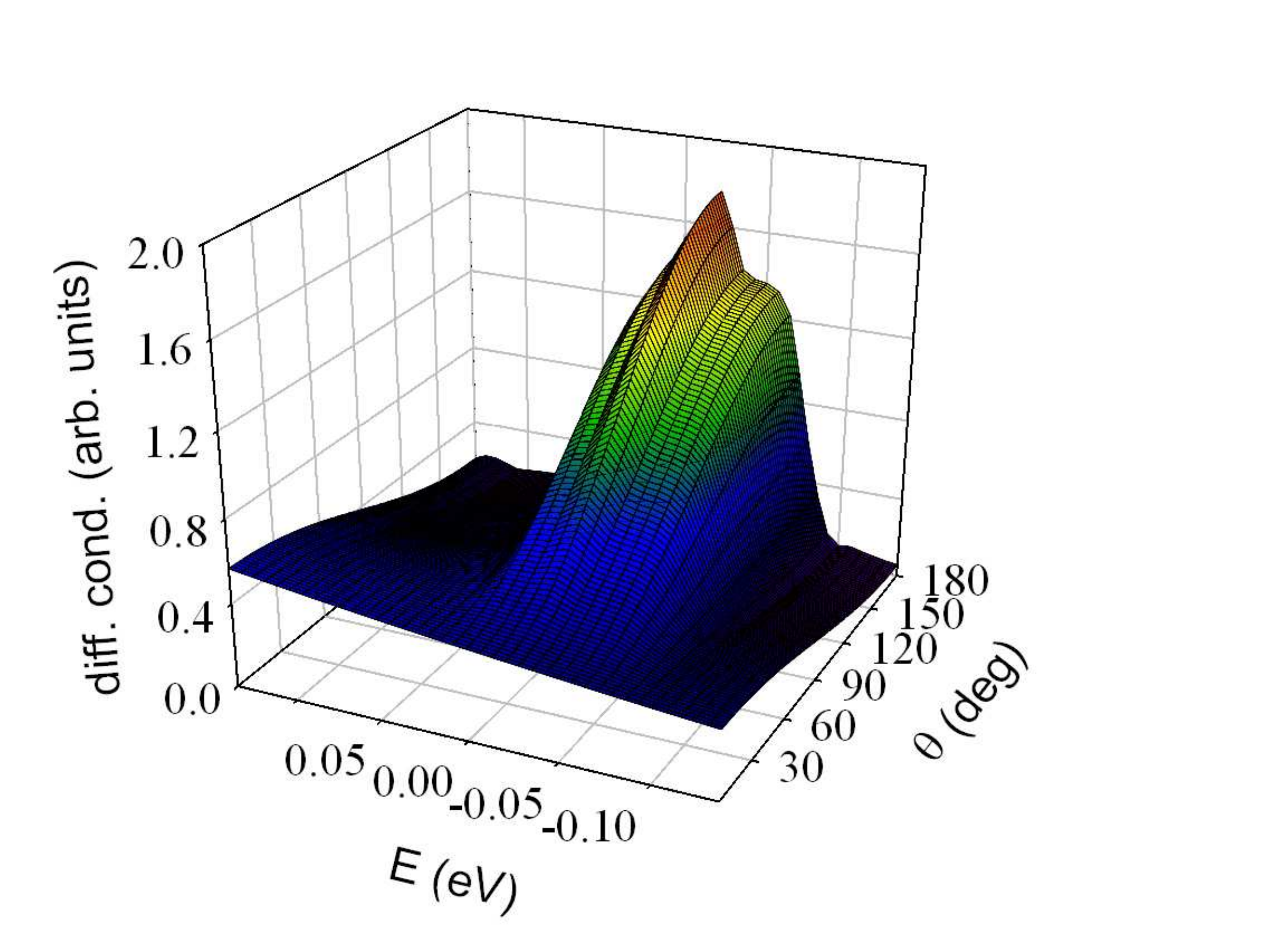}
\caption{This is a three dimensional version of Fig3a) (P=1)} \label{fig4}
\end{figure}

We connect our results from the planar calculation to the realistic experimental STM set-up by the following observation. Following Tersoff and Hamann \cite{tersoff85}, the tunneling 
current at $0$ K is given by
\begin{eqnarray}
&&I(V)\propto\sum_{\bfp\bfk}f\left(E_{i}(\mathbf{p})\right)\left[1-f\left(E_{j}(\mathbf{k})-eV\right)\right]\times\nonumber\\
&&\times|\langle i,\mathbf{p}|T|j,\mathbf{k}\rangle|^2\delta(E_{i}(\mathbf{p})-E_{j}(\mathbf{k}))
\label{eq-I}
\end{eqnarray}
where $f$ is the Fermi function and $|\langle i,\mathbf{p}|T|j,\mathbf{k}\rangle|$
is the tunneling matrix element between initial and final states.
Here, we designate the momentum index $\bfp\ (\bfk)$ to the Cu tip (Fe substrate) electrons and let $i\ (j)$ denote the initial (final) state. We take the convention that electrons flow from the tip to the Fe surface for positive voltage $V$. 
In our planar case, the role of the tip is played by the Cu surface which, with regards to the tunneling, behaves as an STM tip. 
Namely, the electronic structure of Cu is known to be relatively featureless around $E_F$, therefore 
the tunneling matrix element in Eq.~(\ref{eq-I}) only weakly depends on $\mathbf{p}$. The substrate surface, 
on the other hand, has a strong \emph{in-plane} dependence, while the \emph{out-of-plane} component can be written as $\bfk=\bfk_{\|}+q_z$. 
By integrating out the $\mathbf{p}$ dependence we can write (\ref{eq-I}) as
\begin{eqnarray}
&I(V)\propto\sum_{\bfk}\left[1-f\left(E_{j}(\mathbf{k})-eV\right)\right]f(E_{j}(\mathbf{k}))|\langle i|T|j,\mathbf{k}\rangle|^2&\propto\nonumber\\
&\int dE\int d\kpar\left[1-f\left(E_{j}(\mathbf{k})-eV\right)\right]f(E_{j}(\mathbf{k}))|\langle i|T|j,\mathbf{k}\rangle|^2&
\label{eq-II}
\end{eqnarray}
where in the coefficient of proportionality we have incorporated the density of states of the tip (Cu) at $E_F$, and the $\kpar$
integral is taken in the 2DBZ of Fe surface. Equation~(\ref{eq-II}) is equivalent to Eq. (\ref{eq2}), while for a fixed energy
Eq.~(\ref{eq-II}) is equivalent to Eq.~(\ref{eq1}). 
 

Based on these results, we would like to discuss the possibility of detecting the spin-character of Fe surface states with a ferromagnetic (FM) tip. Let's assume that the spin quantization axis in the FM tip is rotated by
an angle $\theta$ relatively to the spin quantization axis in the Fe surface.
The spin components of the electron wave function in the tip can be written
as 
\begin{eqnarray}
&&|\uparrow,\theta \rangle = \cos\left(\theta/2\right)|\uparrow\rangle
-i\sin\left(\theta/2\right)|\downarrow\rangle\nonumber \\
&&|\downarrow,\theta \rangle = -i \sin\left(\theta/2\right) |\uparrow\rangle
+\cos\left(\theta/2\right)|\downarrow\rangle
\label{eq:eq1}
\end{eqnarray}
where, $|\sigma,\theta \rangle$ ($\sigma=\uparrow,\downarrow$) are the spinors for an
arbitrary direction of the spin 
quantization axis and $|\sigma\rangle$ are the spinors for the same spin quantization axis as in Fe.
The total transmission coefficient for an arbitrary spin polarization $P$ 
of the tip can be written as
\begin{widetext}
\begin{eqnarray}
T&=&\left(1+P\right)T^{\uparrow}\cos^{2}\left(\theta/2\right)+ 
  \left(1-P\right)T^{\uparrow}\sin^{2}\left(\theta/2\right)+
  \left(1+P\right)T^{\downarrow}\sin^{2}\left(\theta/2\right)+
  \left(1-P\right)T^{\downarrow}\cos^{2}\left(\theta/2\right)\nonumber\\
&=&\left(T^{\uparrow}+T^{\downarrow}\right)+\left(T^{\uparrow}-T^{\downarrow}\right)P\cos\theta
\end{eqnarray}
\end{widetext}
where, $T^{\uparrow, \downarrow}$ are the spin components of the
transmission coefficient (differential conductance) presented in
Fig.~\ref{Fig1}(c). When $\theta=0$ the spin quantization axis in the
tip is parallel to majority-spins in Fe and when $\theta=180$ it
is parallel to minority-spin.

In Fig.~\ref{Fig3} we show the energy dependence of the differential conductance of a 
FM tip as we rotate the direction of the magnetization in the tip by an angle $\theta$ 
relatively to the magnetization in the Fe. 
Four different 
cases of DOS spin polarization $P$ of the tip are shown.
In the general case, $0<P<1$, a complicated energy-angular dependence is observed.
However, the trends can be easily understood from the two limiting cases of $P=1$ (half-metallic
tip)
and $P=0$ (not shown in Fig.~\ref{Fig3} but can be extrapolated from the case of $P=0.25$).  
For $P=1$ the energy dependence of the transmission for $\theta=0$
is identical to the majority-spin transmission in Fig.~\ref{Fig1}(c), while
for $\theta=180$ the energy dependence is identical to that of minority-spin
in Fig.~\ref{Fig1}(c) (to facilitate the comparison in Fig.~\ref{fig4} we provide 
a three dimensional version of Fig.~\ref{Fig3}a).
As we decrease the degree of spin-polarization in the tip the angular 
dependence of the differential conductance becomes less pronounced.
As it should be, at the limit of $P=0$ the differential conductance becomes independent
of the angle and equal to the sum $T^{\uparrow}+T^{\downarrow}$.
This case corresponds to the conventional (non-magnetic tip) STM measurement 
by Stroscio \emph{ et al}.

\section{CONCLUSION}

In conclusion we have shown that unlike a conventional STM which 
only can measure a sharp peak in the energy dependence of the differential conductance 
when a localized surface band is present, an SP-STM measurement should be able to measure 
an angular dependence as well. 
In the general case of $0<P<1$ a complicated energy-angular dependence emerges but
the trends can be easily understood from the limit of a half-metallic, $P=1$, tip.  
The energy dependence of the differential conductance in this case will be monotonic
for either parallel or anti-parallel direction of the magnetization of the STP tip
relative to the direction of the easy axis in the Fe(001) surface.
This can be used to extract the spin character of the surface band.  
   
\begin{acknowledgments}
The work at Los Alamos National Laboratory
was supported by DOE Office of Basic Energy Sciences Work Proposal
Number 08SCPE973.

\end{acknowledgments}

\bibliography{sstm2}


\end{document}